\documentclass{article}

\usepackage{latexsym}
\usepackage{array}
\usepackage{amsmath}
\usepackage{amssymb}
\usepackage{graphicx}
\usepackage{bm}
\usepackage{color}

\usepackage{theorem}
\theoremstyle{break}
\newtheorem{Theorem}{Theorem}

\def\u{{\upsilon}}

\begin{document}

\title{Singularity confinement and chaos in two-dimensional discrete systems}
\author{Masataka Kanki$^1$, Takafumi Mase$^2$ and
Tetsuji Tokihiro$^2$\\
\small $^1$ Faculty of Engineering Science,\\
\small Kansai University, 3-3-35 Yamate, Suita, Osaka 564-8680, Japan\\
\small $^2$ Graduate School of Mathematical Sciences,\\
\small University of Tokyo, 3-8-1 Komaba, Tokyo 153-8914, Japan}

\date{}

\maketitle

\begin{abstract}
We present a quasi-integrable two-dimensional lattice equation: i.e., a partial difference equation which satisfies a test for integrability, singularity confinement, although it has a chaotic aspect in the sense that the degrees of its iterates exhibit exponential growth. By systematic reduction to one-dimensional systems, it gives a hierarchy of ordinary difference equations with confined singularities, but with positive algebraic entropy including a generalized form of the Hietarinta-Viallet mapping. We believe that this is the first example of such quasi-integrable equations defined over a two-dimensional lattice.

\end{abstract}

\section{Introduction}
Dynamical systems can be classified into two broad categories; integrable systems and non-integrable systems.
For Hamiltonian equations of motion with finite degrees of freedom, an integrable system is defined to have enough number (half the degrees of freedom) of independent constants of motion in involution which allow us to obtain local and global information about the system explicitly or implicitly\cite{Arnold}.
This integrability criterion is extended to nonlinear partial differential equations and the equations with this property are called infinite dimensional non-linear integrable equations or soliton equations\cite{MJD, Ablowitz}.
Another test for integrability is the Painlev\'{e} property which is based on singularity analysis of solutions\cite{Conte, ARS}.
The Painlev\'{e} equations are the second order differential equations with this property, and they play important roles in many areas of physics, e.g. in the theory of correlation functions of quantum spin systems, random matrices and so on\cite{Wu, TracyWidom}.  
For discrete dynamical systems, an analog of the Painlev\'{e} property is the singularity confinement\cite{SC}, by which a series of discrete Painlev\'{e} equations were successfully constructed\cite{RGH}.
The QRT mappings, which are typical second order integrable mappings with a constant of motion, also satisfy singularity confinement test\cite{QRT}.

We have to note that there are second order mappings which have singularity confinement but show chaotic behaviors, and therefore singularity confinement is only a necessary property for the mapping to be integrable.
The first and one of the most important examples was given by Hietarinta and Viallet \cite{HV}. 
They introduced the following second order recurrence relation
\begin{equation} \label{HVeq}
x_{n+1}=-x_{n-1}+x_n+\frac{a}{x_n^2}.
\end{equation}
If $x_n$ reaches $0$, then we have $x_{n+1}=\infty$, $x_{n+2}=\infty$ and we encounter indeterminacy $x_{n+3}=\infty -\infty$.
However, if we use an infinitesimally small parameter $\epsilon$ and set $x_{n-1}=c,\,x_n=\epsilon$, we have $x_{n+1}=O(\epsilon^{-2})$,
$x_{n+2}=O(\epsilon^{-2})$, $x_{n+3}=O(\epsilon)$ and $x_{n+4}=c+O(\epsilon)$.
By letting $\epsilon \to 0$, we have a definite sequence $...,c,0,\infty,\infty,0,c,c+a/c^2,...$, hence the singularity is confined.
But the orbits of \eqref{HVeq} are shown to exhibit chaotic behaviors \cite{HV}.
To refine the integrability test, the increasing rate of the degree of the iterates was investigates in \cite{HV}.
The complexity of discrete systems are measured by the degrees of the successive iterates $\phi^n$ of the mapping $\phi$.
When the degree $\deg \phi=d$, we na\"{i}vely assume that $\deg \phi^n\sim d^n$, which is an exponential growth, however, for mappings with strong cancellations during the iterations, $\deg \phi^n$ may grow only polynomially. The algebraic entropy \cite{BV} of the mapping $\phi$ is defined by
\[
\lambda:=\lim_{n\to \infty} \frac{1}{n}\ln (\deg \phi^n),
\]
which is always convergent to a non-negative value.
Algebraic entropy is a more sensitive integrability test than singularity confinement,
and the dynamical system is, in most cases, integrable if its algebraic entropy is zero \cite{Veselov}.
In other words, if the degrees of the iterates grows faster than polynomially, it is likely that the system is non-integrable.
In \cite{HV}, it is proved that the algebraic entropy of \eqref{HVeq} is positive ($\ln((3+\sqrt{5})/2)=0.962...$), and from this observation, they have proposed that the algebraic entropy can be a refined integrability detector.
%%%%%%%%%%%%%%%%%%%%%%%%%%%%%%%%%%%%%%%%%%%%%%%%%%%%%%
%
%%%%%%%%%%%%%%%%%%%%%%%%%%%%%%%%%%%%%%%%%%%%%%%%%%%%%%%

Hereafter we will use the new term {\it quasi-integrable} to denote the discrete systems which have confined singularities and the positive algebraic entropy at the same time.
Hietarinta-Viallet equation \eqref{HVeq} is quasi-integrable and subsequently many quasi-integrable second order mappings were discovered\cite{Tsuda, Bedford2, Bedford, Redemption, Mase2015}.
A natural expectation would be the existence of a higher dimensional quasi-integrable system.
However, up to our best knowledge, neither two-dimensional quasi-integrable equation nor higher than second order quasi-integrable mapping has been reported yet.
A reason why a higher dimensional quasi-integrable mapping was not found for a long time is that the number of singularity patterns of the system becomes infinite and it is practically impossible to check all of them, though some of the discrete soliton equations were discussed in view of confined singularities\cite{RGS}.

Recently an algebraic reinterpretation of singularity confinement, {\it co-primeness}, was proposed so that it can apply to higher dimensional systems\cite{dKdVSC, dKdVSC2}. 
For a second order rational mapping, $(x_n,x_{n-1}) \mapsto (x_{n+1},x_n)$, $x_n$ is regarded as a rational function of the initial data $(x_0,x_1)$. 
We say that the mapping has {\it co-primeness} if there exists a positive integer $M$ such that any pair of $x_n$ and $x_m$ with $|n-m|>M$ has no common factor except for a monomial of $x_0,\,x_1$. 
For example, in mapping \eqref{HVeq}, we can prove that $x_n=p_{n+2}p_{n-1}/(p_{n+1}p_n)^2$ where $p_n$ is an independent {\it irreducible} Laurent polynomial of $x_0,x_1$, and that $x_n$ and $x_m$ have no common factor except for a monomial of $x_0,\,x_1$ on condition that $|n-m|>3$ \cite{extHV}. 
The notion of co-primeness has been introduced as an alternative for the singularity confinement test that works better for two-dimensional lattice equations, and  the discrete KdV equation and the discrete Toda equations have been shown to have co-primeness for various boundary conditions\cite{dKdVSC2, dToda}.
On the other hand, non-confining systems do not have co-primeness. For example, a non-confining and non-integrable mapping $x_{n+1}=(x_n+1)/(x_{n-1}x_n^3)$ does not have co-primeness, since $x_n$ and $x_m$ share the factor $(x_1+1)$ for all $n, m\ge 2$.
Another example of mapping without co-primeness property is the following mapping
\begin{equation}
x_{n+1}=-x_{n-1}+x_n+\frac{1}{x_n^k},\ \ (k\ge 3,\ k\ \text{is odd})
\end{equation}
which has been proposed as a non-confining generalization of the Hietarinta-Viallet equation in \cite{extHV}.

In this Letter, utilizing the notion of co-primeness, we will present one example of two-dimensional quasi-integrable lattice equation (equation \eqref{eq11}).
Moreover, equation \eqref{eq11} is shown to give a hierarchy of quasi-integrable one-dimensional systems, including the Hietarinta-Viallet equation \eqref{HVeq}, by projecting from the two-dimensional lattice.
We also study the algebraic entropy of the four-term (third-order) recurrence relation corresponding to \eqref{HVeq}, which could not have been investigated with conventional approach using an algebraic geometry of rational varieties.
%%%%%%%%%%%%%%%%%%%%%%%%%%%%%%%%%%%%%%%%%%%%%%%%%%%%%%
%
%%%%%%%%%%%%%%%%%%%%%%%%%%%%%%%%%%%%%%%%%%%%%%%%%%%%%%%

\section{Quasi-integrable lattice equation}
\begin{figure}[h]
  \begin{center}
    \includegraphics[bb=200 600 400 750,width=7.0cm]{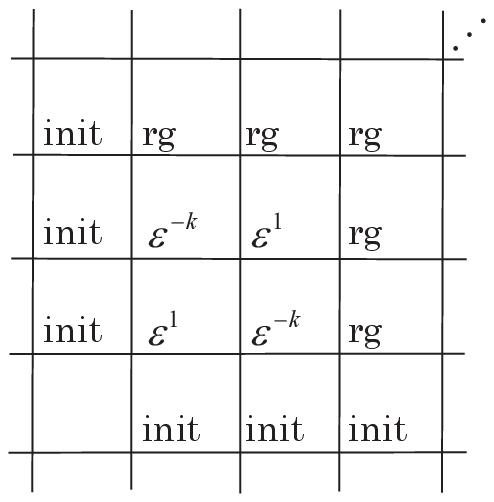}
    \caption{Singularity pattern for the lattice equation \eqref{eq11}.
Here, `init' denotes a point of initial value, and `rg' denotes that of regular value.}
    \label{fig1}
  \end{center}
\end{figure}

Let $k\ge 2$ be an even integer.
Our main target is the following partial difference equation
\begin{equation}
x_{t,n}= - x_{t-1,n-1}+\frac{a}{x_{t,n-1}^k}+\frac{b}{x_{t-1,n}^k},\ \ (k\ \text{is even}\ ) \label{eq11}
\end{equation}
where $a,b \neq 0$ are constants. Equation \eqref{eq11} is a lattice equation over $(t,n)\in \mathbb{Z}^2$.
Under a suitable set of initial data, time evolution of \eqref{eq11} is well-defined toward an upper right of the $n$-$t$ plane.
In case $k=1$ and $a=b$, \eqref{eq11} is equivalent to the discrete KdV equation and we may regard \eqref{eq11} as a quasi-integrable generalization of the discrete KdV equation. 
Let us take initial data of \eqref{eq11} as $x_{1,1}=\varepsilon$, where $\varepsilon$ is an infinitesimal.
Then the singularity pattern of \eqref{eq11} over the $t$-$n$ lattice is obtained as in Fig. \ref{fig1}, where the lower left corner indicates a point $(t,n)=(0,0)$, and we have placed the evaluation of each cell on the lattice.
Here, the evolution of \eqref{eq11} is defined from $x_{1,1}=\varepsilon$ and generic initial values $x_{1,0}$, $x_{0,1}$, $x_{2,0}$, $x_{0,2}$, $x_{3,0}$, $x_{0,3}$.
By ``rg'', we denote that the iterate is regular valued (i.e., $x_{t,n}$ is not $\infty$ and retains the dependence on the initial data).
Also, we use an expression $x \sim \varepsilon^m$ to indicate that the highest order of the $\varepsilon$-expansion of $x$ is $\varepsilon^m$.

Let us briefly explain how to prove the confinement.
First $x_{2,1}=-x_{1,0}+a x_{2,0}^
{-k} +b \varepsilon^{-k}\sim \varepsilon^{-k}$ and $x_{1,2}\sim \varepsilon^{-k}$.
Using these values we have $x_{2,2}=-\varepsilon+O(\varepsilon^{k^2})\sim \varepsilon$, $x_{3,1}=-x_{2,0}+a x_{3,0}^{-k}+O(\varepsilon^{k^2})\sim \text{rg}$, and $x_{1,3}\sim \text{rg}$. At the next iterate $x_{3,2}$, since
\begin{align}
x_{3,2}&=-\frac{b}{\varepsilon^k}+x_{1,0}-\frac{a}{x_{2,0}^k} \notag \\
& + a\left( O(1) \right)^{-k}+b\left(-\varepsilon+O(\varepsilon^{k^2})\right)^{-k}, \notag
\end{align}
\[
 b\left(-\varepsilon+O(\varepsilon^{k^2})\right)^{-k}=(-1)^k \frac{b}{\varepsilon^k}+O(\varepsilon^{k^2-k-1})
\]
we have a non-trivial cancellation when $k$ is even, and $x_{3,2}\sim \text{rg}$.
In the same manner, $x_{2,3},x_{3,3}$ are regular valued, which conclude that \eqref{eq11} passes the singularity confinement test for this fundamental singularity pattern.
Note that the singularity is not confined when $k(\ge 3)$ is an odd integer.

To show that singularities are confined in general, we prove that \eqref{eq11} has co-primeness.
For this purpose, let us introduce another variable $\upsilon_{t,n}$, $((t,n)\in\mathbb{Z}^2)$ by the relation \eqref{transform11}
\begin{equation}
x_{t,n}=\frac{\upsilon_{t,n} \upsilon_{t-1,n-1}}{\upsilon_{t-1,n}^k \upsilon_{t,n-1}^k}, \label{transform11}
\end{equation}
which is inferred from the singularity pattern of $x_{1,1},x_{2,1},x_{1,2},x_{2,2}$ above. We obtain the following higher order recurrence relation
\begin{align}
\u_{t,n}&=\frac{1}{\u_{t-2,n-1}^k \u_{t-1,n-2}^k}\Big( -\u_{t-2,n-2} \u_{t-1,n}^k \u_{t,n-1}^k \notag \\
& \quad +a \u_{t-1,n-1}^{k^2-1}\u_{t,n-2}^{k^2}\u_{t-1,n}^k \u_{t-2,n-1}^k\notag \\
& \qquad +b \u_{t-1,n-1}^{k^2-1} \u_{t-2,n}^{k^2} \u_{t,n-1}^k \u_{t-1,n-2}^k \Big). \label{eq22}
\end{align}
Although \eqref{eq22} is not a multi-linear equation, it helps us in a way similar to bilinear forms of integrable equations.
The equation satisfies the Laurent property, the irreducibility property and the co-primeness property, which tell us that \eqref{eq22} and thus \eqref{eq11} is quasi-integrable. Let $a,b\in\mathbb{C}^{\times}$, and let $H:=\{(t,n)\in\mathbb{Z}^2\, |\, t \in\{0,1\}\ \mbox{or}\ n\in\{0,1\}\}$ be the domain on which the initial values for \eqref{eq22} sit.
Then the iterates $\u_{t,n}$ on the first quadrant $(t\ge 0, n\ge 0)$ are well-defined from \eqref{eq22} and satisfy
\[
\u_{t,n}\in \mathbb{C}[\u_{s,m},\u_{s,m}^{-1}\, | \, (s,m)\in H],
\]
that is, $\u_{t,n}$ is a Laurent polynomial of $\{\u_{s,m}\}_{(s,m)\in H}$.
Moreover, each iterate $\u_{t,n}$ is an independent irreducible Laurent polynomial.
The proof of this statement is done by induction in a similar way to the proofs shown in Ref.~\cite{dKdVSC2, dToda}, but it is fairly technical and will be reported elsewhere.
For example, when $k=2$, at $\u_{3,4}$, we have a non-trivial cancellation of $\u_{2,2}^2$ and only the monomials of initial variables remain in the denominator.

From these results, we can obtain the co-primeness property for the partial difference equation \eqref{eq11}.
We state this fact as a theorem:
\begin{Theorem} \label{coprimethm}
Let $k\ge 2$ be an even integer and let $G:=\{(t,n)\in\mathbb{Z}^2\, |\, t=0\ \mbox{or}\ n=0\}$ be the domain of initial values of \eqref{eq11}.
Every pair of iterates $x_{t,n}$ and $x_{s,m}$ of \eqref{eq11} is co-prime in the field $\mathbb{C}(x_{t,n}|(t,n)\in G)$ of rational functions, if it satisfies at least one of $|t-s|>1$ or $|n-m|>1$.
\end{Theorem}
Here two rational functions $F_1/F_2$ and $P_1/P_2$ are co-prime if no pair among $(F_1,P_1)$, $(F_1,P_2)$, $(F_2,P_1)$, $(F_2,P_2)$ has common factors except for the monomials of initial variables.
The proof of theorem \ref{coprimethm} is done by giving a direct correspondence between the initial values of  \eqref{eq11} and those of \eqref{eq22}.
For example,
\[x_{t,n}=(\upsilon_{t,n} \upsilon_{t-1,n-1})/(\upsilon_{t-1,n}^k \upsilon_{t,n-1}^k)\]
and
\[x_{t-2,n-2}=(\upsilon_{t-2,n-2} \upsilon_{t-3,n-3})/ (\upsilon_{t-3,n-2}^k \upsilon_{t-2,n-3}^k)\]
do not share a common factor, since each $\u_{m,s}$ is irreducible.

We can generalize equation \eqref{eq11}:
\begin{equation}
x_{t,n}= -c_{t,n} x_{t-1,n-1}+\frac{a_{t,n}}{x_{t,n-1}^k}+\frac{b_{t,n}}{x_{t-1,n}^m}, \label{eq11gen}
\end{equation}
where $a_{t,n},b_{t,n},c_{t,n}$ are non-zero for every $t$ and $n$, and $k,m$ are integers greater than one.
If we assume that the singularity is confined in the same manner as that for equation \eqref{eq11},
we obtain the following two conditions for non-autonomous terms:\[
c_{t+1,n+1} c_{t,n+1}^m b_{t,n}=(-1)^m b_{t+1,n+1},\ \text{and}\ c_{t+1,n+1} c_{t+1,n}^k a_{t,n}=(-1)^k a_{t+1,n+1}.
\]
As special cases, we just present two examples:
\begin{equation}
x_{t,n}= - x_{t-1,n-1}+\frac{a}{x_{t,n-1}^k}+\frac{b}{x_{t-1,n}^m}\ (k,m\ \text{are both even}),
\end{equation}
which is the only case of autonomous equation, and
\begin{equation}
x_{t,n}= x_{t-1,n-1}+\frac{a_{t,n}}{x_{t,n-1}^k}+\frac{b_{t,n}}{x_{t-1,n}^m},
\end{equation}
where $c_{t,n}=-1$, $a_{t+1,n+1}=-a_{t,n}$ and $b_{t+1,n+1}=-b_{t,n}$ for every $t,n$.
Again we obtain a generalized form of the discrete KdV equation.

%%%%%%%%%%%%%%%%%%%%%%%%%%%%%%%%%%%%%%%%%%%%%%%%%%%%%%
%
%%%%%%%%%%%%%%%%%%%%%%%%%%%%%%%%%%%%%%%%%%%%%%%%%%%%%%%

\section{Reduction}
Now we examine a relation of \eqref{eq11} with one-dimensional discrete systems such as Hietarinta-Viallet equation \cite{HV} and its extension \cite{extHV}. Let $p,q$ be positive integers.
Reduction of \eqref{eq11} is done by identifying those iterates $x_{t,n}$ that have the same value for $pt+qn$.
By introducing a new one-dimensional variable $y_m:=x_{t,n}$ for $m=pt+qn$, we obtain the following reduced system
\begin{equation}\label{reducedeq11}
y_m=-y_{m-p-q}+\frac{a}{y_{m-q}^k}+\frac{b}{y_{m-p}^k}.
\end{equation}
The transformation \eqref{transform11} and the $\u$-functions \eqref{eq22} are reduced in the same manner: i.e., \eqref{eq22} is reduced to the following nonlinear form:
\begin{align}
u_m&=\frac{1}{u_{m-2p-q}^k u_{m-p-2q}^k}\Big( -u_{m-2p-2q} u_{m-p}^k u_{m-q}^k \notag\\ 
&\quad+a u_{m-p-q}^{k^2-1} u_{m-2q}^{k^2} u_{m-p}^k u_{m-2p-q}^k\notag \\
& \qquad +b u_{m-p-q}^{k^2-1} u_{m-2p}^{k^2} u_{m-q}^k u_{m-p-2q}^k \Big), \label{reducedeq22}
\end{align}
by the transformation
\begin{equation} \label{ymfactor}
y_m=\frac{u_m u_{m-p-q}}{u_{m-p}^k u_{m-q}^k}.
\end{equation}
It is not hard to prove that every iterate $u_m$ of \eqref{reducedeq22} satisfies the Laurent property: $u_m\in\mathbb{C}[u_k,u_k^{-1}\, |\, 0\le k\le 2p+2q-1]$, utilizing the Laurent property for equation \eqref{eq22}.
We use the fact that the Laurent property is also satisfied for the domain of initial values $\{(t,n)\, |\, pt+qn=0,1,\cdots,p+q-1\}$, which is different from $G$.
Therefore, identifying the iterates $\u_{t,n}$ with fixed $pt+qn$ does not break the Laurent property of $\u_{t,n}$.

On the other hand, the irreducibility and the co-primeness theorem \ref{coprimethm} is not trivially reduced to these properties for \eqref{reducedeq11} and \eqref{reducedeq22}. We have proved that the irreducibility and the co-primeness hold for $(p,q)=(1,2)$. However, for larger $p,q$, we just give two conjectures that, if $p\neq q$, (i) every iterate $u_m$ is an {\it irreducible} Laurent polynomial, and that (ii) two iterates $x_m$ and $x_l$ are co-prime
in the field of rational functions $\mathbb{C}(x_k\, |\, 0\le k\le p+q-1)$ on condition that $|m-l|>2p+2q$, which is immediate from (i).
Moreover, the algebraic entropy of equation \eqref{reducedeq11} is conjectured to be $\ln \Lambda>0$, where $\Lambda$ is the largest real root of 
\begin{equation}
1-k \Lambda^p -k \Lambda^q+\Lambda^{p+q}=0. \label{ymalgent}
\end{equation}
If we suppose the irreducibility of $u_m$, the factorization form of $y_m$ in \eqref{ymfactor} does not have additional cancellations. Thus, if we suppose the increasing rate as $u_m\sim \Lambda^m$, we obtain the desired polynomial \eqref{ymalgent}.

Equation \eqref{reducedeq11} includes an extended form of the Hietarinta-Viallet equation which is seen by a simple {\it integration} procedure for recurrence relations.
Let us take $p=1$ and $q=2$ and suppose that $a=b$ in \eqref{reducedeq11}: i.e., we consider
\begin{equation}
y_m=-y_{m-3}+\frac{a}{y_{m-2}^k}+\frac{a}{y_{m-1}^k}. \label{HVk4term}
\end{equation}
Since \eqref{HVk4term} implies
\begin{align*}
&y_{m+1}-y_m-\frac{a}{y_m^k}+y_{m-1}\\
&=-\left(y_m-y_{m-1}-\frac{a}{y_{m-1}^k}+y_{m-2} \right),
\end{align*}
we can integrate
\[
y_{m+1}-y_m-\frac{a}{y_m^k}+y_{m-1}=(-1)^m C,
\]
where $C$ is a constant determined by the initial condition.
In particular, if we take the initial condition as $C=0$, we obtain
\begin{equation}
y_{m+1}=-y_{m-1}+y_m+\frac{a}{y_m^k}, \label{HVk}
\end{equation}
which is nothing but the extended Hietarinta-Viallet equation we have investigated in \cite{extHV}. 
Equation of the type \eqref{HVk} is studied from algebro-geometric approach \cite{Bedford}.
For $k=2$ we recover the original chaotic equation by Hietarinta and Viallet \cite{HV}.
Equation \eqref{HVk} is shown to pass the singularity confinement test and to have the irreducibility and co-primeness property if and only if $k>1$ is an even integer. It is also known that the algebraic entropy of \eqref{HVk} for $k$ even is $\lambda_k:=\ln\{(k+1+\sqrt{(k-1)(k+3)})/2\}>0$. Therefore \eqref{HVk} is a quasi-integrable discrete system.
From equation \eqref{ymalgent}, the $4$-term recurrence relation \eqref{HVk4term} seems to have the same algebraic entropy $\lambda_k$ as that of \eqref{HVk}.

We present some numerical calculation based on the notion of Diophantine integrability \cite{Halburd}.
The idea of Diophantine integrability is to take the initial values and the coefficients as rational numbers so that all the iterates $y_m (m\ge 3)$ are rational valued. We use the height of rational numbers instead of the degrees of the iterates to define {\it Diophantine entropy}. For a rational number $r/s\in\mathbb{Q}$ which is an irreducible fraction, the height $H(r/s)$ is defined as $H(r/s)=\max\{|r|,|s|\}$. For an arbitrary discrete dynamical systems with rational coefficients, let us fix the initial values $y_0,y_1,y_2,...\in\mathbb{Q}$ and numerically compute
$M_m:=(\ln H(y_{m+1}))/(\ln H(y_m))$.
$M_m$ is conjectured to converge to the exponential of the algebraic entropy, for generic initial conditions.
For example, in equation \eqref{HVk4term}, if we take $a=1$ and the initial values as $y_0=2$,  $y_1=1$, $y_2=2$, then $M_{18}=2.6180...$ for $k=2$, $M_{12}= 4.7912...$ for $k=4$, $M_{10}= 6.8541...$ for $k=6$, which show good convergence to $\exp(\lambda_k)$ even for small $m$.
Proof is an open problem, however, we can at least prove that the lower bound for the algebraic entropy of \eqref{HVk4term} is at least $\lambda_k$.
Thus the equation \eqref{HVk4term} is a chaotic equation with a positive entropy.

%%%%%%%%%%%%%%%%%%%%%%%%%%%%%%%%%%%%%%%%%%%%%%%%%%%%%%
%
%%%%%%%%%%%%%%%%%%%%%%%%%%%%%%%%%%%%%%%%%%%%%%%%%%%%%%%

\section{Conclusion}
We have presented a novel example of two-dimensional {\it quasi-integrable} discrete dynamical system \eqref{eq11}, with confined singularities, but with exponential growth of degrees.
The system we have introduced can be reduced to a hierarchy of quasi-integrable one-dimensional equations of order higher than two (equation \eqref{reducedeq11}),
which pass the singularity confinement test and the co-primeness criterion, but nevertheless are
non-integrable with positive entropy.
The exact value of the algebraic entropy is conjectured, by comparing the degrees of the denominator and the numerator of the iterates. The discussion relies on the irreducibility and the co-primeness property, which has been proposed as a new necessary condition for integrability.
Our one-dimensional systems include the celebrated Hietarinta-Viallet equation, which is the first example reported to have confined singularities and at the same time is non-integrable with orbits of chaotic behavior.

Chaos is a universal phenomenon appearing in nonlinear dynamical systems and has wide application to science and engineering \cite{Strogatz}. A quasi-integrable system, which has an aspect of integrability, belongs to a particular class of chaotic systems,
and will have potential application to complex phenomena, by enabling us to avoid singularities in numerical simulations. Furthermore, by utilizing the co-primeness and the irreducibility, it becomes tractable for us to estimate physical quantities related to the initial data in quasi-integrable systems. The two-dimensional quasi-integrable systems we have presented here suggest that, there exists a class of quasi-integrable systems just like that of integrable ones, which will be of considerable importance in the study of higher dimensional nonlinear systems.
To find and to investigate wider classes of quasi-integrable systems, and to apply them to various fields of science are the subjects we wish to address in future.

\section*{Acknowledgments}
The authors thank Prof. R. Willox for useful comments.
This work is partially supported by KAKENHI Grant Numbers 15H06128, 25-3088 and the Program for Leading Graduate Schools, MEXT, Japan.
%%%%%%%%%%%%%%%%%%%%%%%%%%%%%
%

\end{document}